\documentclass[5p,times]{elsarticle}
\DeclareUnicodeCharacter{25CF}{\textbullet}

\usepackage{colortbl}            %
\usepackage{hhline}              %
\usepackage{graphicx}            %
\usepackage{xurl}                %

\usepackage{hyperref}
\usepackage{etoolbox}
\appto\UrlBreaks{\do\-\do\-\do\.}\usepackage{graphicx}
\usepackage[colorinlistoftodos]{todonotes}
\usepackage{array}
\usepackage{wrapfig}
\usepackage{multirow}
\usepackage{tabularx}
\usepackage{orcidlink}
\usepackage[svgnames,table]{xcolor}

\usepackage{multirow}
\usepackage{booktabs}
\usepackage{adjustbox}

\usepackage[T1]{fontenc}
\usepackage[utf8]{inputenc}
\usepackage[colorinlistoftodos]{todonotes}
\usepackage{soul}

\definecolor{headerGray}{HTML}{F2F2F2} %
\definecolor{ruleGray}{HTML}{DDDDDD}   %

\newcolumntype{L}{>{\raggedright\arraybackslash}X}
\newcolumntype{P}[1]{>{\raggedright\arraybackslash}p{#1}}

\hypersetup{
    colorlinks=false, %
    linkcolor=blue,  %
    citecolor=green, %
    filecolor=magenta, %
    urlcolor=black,   %
    breaklinks=true  %
}

\colorlet{myblue}{blue!10!white}
\colorlet{myorange}{orange!10!white}
\colorlet{mygreen}{green!10!white}

\makeatletter
 \if@todonotes@disabled
 
 \else
 
 \fi
 \makeatother

\makeatletter
 \if@todonotes@disabled
 
 \else
 
 \fi
 \makeatother\newsavebox{\selvestebox}

\makeatletter
 \if@todonotes@disabled
 
 \else
 
 \fi
 \makeatother

\begin{document}
\begin{frontmatter}

\title{Integrated Data, Metadata, and Paradata Management System for 3D Digital Cultural Heritage Objects: Workflow Automation, Federated Authentication, and Publication}\author[inst1,inst2]{Micha\l{} Orzechowski\orcidlink{0000-0002-8558-1283}}
\author[inst1]{\L{}ukasz Opio\l{}a\orcidlink{0000-0002-5981-1727}}
\author[inst4]{Ignacio Lamata Martínez\orcidlink{0000-0002-5981-1727}}
\author[inst3]{Marinos Ioannides\orcidlink{0009-0007-8181-7926}}
\author[inst3]{Panayiotis N. Panayiotou\orcidlink{0009-0007-8181-7926}}
\author[inst1]{\L{}ukasz Dutka\orcidlink{0000-0002-8781-866X}}
\author[inst2]{Renata G. S\l{}ota\orcidlink{0000-0001-7424-9317}}
\author[inst1,inst2]{Jacek Kitowski\orcidlink{0000-0003-3902-8310}}

\affiliation[inst1]{organization={Academic Computer Centre CYFRONET AGH},
            addressline={Nawojki 11}, 
            city={Krakow},
            country={Poland}}
            
\affiliation[inst2]{organization={AGH University of Krakow},
            addressline={Adama Mickiewicza 30}, 
            city={Krakow},
            country={Poland}}

\affiliation[inst3]{organization={UNESCO Chair on Digital Cultural Heritage at Cyprus University of Technology},
            addressline={Research Centre MNEMOSYNE}, 
            city={Limassol},
            country={Cyprus}}

\affiliation[inst4]{organization={EGI Foundation},
            addressline={Science Park 140}, 
            city={Amsterdam},
            country={The Netherlands}}

\begin{abstract}

The complexity of high-quality 3D digitised cultural heritage objects creates challenges for existing data management systems as they need to develop metadata management and processing capabilities to provide semantic insight into the interconnectivity of data that constitutes cultural heritage objects. 
To address these challenges, we propose a global federated authentication and authorisation mechanism, a data and metadata management system, and an integrated engine for designing and executing automated workflows that facilitate the processing of both data and metadata. The solution is evaluated with three distinct 3D digitised cultural objects and presents the complete process from data upload to cultural heritage object publication.\end{abstract}
\end{frontmatter}

\section{Introduction}\label{section:introduction}

Over the last two decades, the European Union has allocated significant resources~\cite{ch-investiment-1}\cite{ch-investiment-2}\cite{ch-investiment-3} towards the digitalisation of Cultural Heritage (CH) assets. This funding 
has been instrumental in supporting CH entities like museums, archives, and libraries in digitising, archiving, and preserving their collections for research, innovation, and educational purposes. This investment period has been pivotal for the multi/interdisciplinary community and cross-sectoral development in the European CH sector. 
In parallel, the last years have witnessed significant private, governmental, and non-governmental initiatives in 2D and 3D digitalisation from entities such as Google~\cite{google-heritage}, CyArk~\cite{cyark}, or the Smithsonian Institute~\cite{smithsonian-institute}. The advancements in Cultural Heritage Institutions (CHIs) have catalysed an exponential increase in digital content production, with the digital universe expected to expand tenfold in the coming years. Significant strides in 3D digitalisation have enhanced the accessibility of European CH for multiple purposes. High-quality 3D scans aid in archaeological and engineering conservation efforts, while medium-quality data finds applications in the creative industry. 

CHIs encounter several obstacles when disseminating 3D models to interdisciplinary audiences.  The file formats commonly used for storing 3D data and associated raw data may be unsuitable for online visualization and delivery. This is primarily due to large file sizes and the significant client-side processing required to render 3D content. Also, although there exist established standards for storing and processing 3D objects' metadata, this is not yet true about the paradata as information describing the process of digitisation of the CHO and the provenance of the 3D object.
While these challenges also affect 2D content, they are more prominent in the case of 3D due to the inherent complexity and substantially larger storage requirements of 3D models, which impact network transfer speeds and processing capabilities. To mitigate these issues, CHIs can employ strategies such as providing multiple versions of a 3D model optimized for different use cases. This allows them to cater to the specific needs of multidisciplinary users, based on the CHI's digitisation use case reusability assessment, via various channels, including proprietary or open-access platforms, self-hosted or cloud infrastructures.
The EU-funded Europeana Foundation~\cite{europeana} serves as a major centralized digital repository and dissemination platform for European cultural heritage, aggregating and providing access to digitized materials from various CHIs, such as museums and libraries. However, CHIs face significant challenges in sharing 3D models within the Europeana platform, including:
\begin{itemize}
\item heterogeneous storage and access requirements: CHIs must accommodate diverse 3D model formats and provide tailored access to various user groups with differing needs and technical capabilities.
\item online visualization: Effective online visualization of 3D models within Europeana is crucial for both discovery and access, demanding efficient rendering and streaming solutions.
\item metadata management: CHIs need to create and maintain standardized metadata for their 3D models, ensuring alignment with the evolving Europeana Data Model~\cite{europeana2025edm} (EDM),
\item paradata availability: sharing comprehensive paradata, documenting the entire 2D/3D data acquisition and modelling process.
\end{itemize}
Currently, CHIs often address these challenges in an ad hoc manner, relying on disparate in-house or outsourced services. This fragmented approach leads to: \begin{itemize}
\item duplication of effort: multiple CHIs may independently develop similar solutions or workflows for 3D model management,
\item redundancies: data and metadata may be stored redundantly across different systems,
\item complex process management: coordinating multiple service providers and integrating diverse tools and technologies creates complex and inefficient processes.
\end{itemize}

This situation highlights the need for standardized practices, tools, and infrastructure to support CHIs in effectively sharing 3D models within Europeana.
In this paper, we present our solution for 3D Cultural Heritage Objects (CHO) management, processing, and integration with external cultural heritage digital collections.
The proposed implementation combines data access and management functionality, metadata management, processing capabilities of the integrated workflow automation platform, the federated authentication and authorisation mechanism, and finally, the ability to publish the resulting CHO to Europeana. 
The novelty of this paper comprises:
\begin{itemize}

\item integration of metadata and paradata management of CHO with a data management platform,

\item introduction of automated workflows for processing of CHO's data, metadata, and paradata,

\item implementation of EDM as a part of the data management platform to facilitate CHO management and publication,

\end{itemize}
The solution is built using existing data management~\cite{onedata-wrzeszcz-fgcs} and federated authentication~\cite{egi-checkin} tools, extending them with new capabilities:  support for distinctive metadata and paradata of CHO, a dedicated EDM editor to manage and process the CHO-specific metadata and paradata, specialized CHO management workflows, and an integrated publication mechanism of CHO to Europeana.

The paper is organised as follows. Section~\ref{section:relatedwork} discusses the problem statement and related work in the area of digital infrastructures of management and processing of 3D objects. Section \ref{section:onedata} describes the integrated automation workflow implementation as part of the Onedata platform. Section~\ref{section:CHO-solution} presents our solution for CHO processing, in particular the federated authentication and authorisation mechanism, data metadata processing, and CHO publication. 
Section~\ref{section:evaluation} provides an evaluation of the solution, describing the comprehensive processing of a complex 3D cultural heritage objects. Finally, Section~\ref{section:conclusions} concludes the paper. This paper is an extension of the conference work \cite{ppam2024}, where we describe the initial version of the 3D Cultural Heritage Objects (CHO) management, processing, and integration with external cultural heritage digital collections.

\section{Problem Statement and Related Work}\label{section:relatedwork}
The 3D digitalisation of movable and immovable CHO is an inherently complex, multi-stage process typically undertaken by professionals from companies specializing in 3D modelling, such as 3D model makers and graphic designers, digital cultural researchers, scientific project managers, and engineers working at technology universities. Different parameters, such as infrastructure hardware, software, personnel, etc., could influence this process and yield different results. Furthermore, there are unique documentation challenges for the CHOs within each organisational category, and the capabilities of the recording hardware, associated processing software, production methods, and online visualisation systems are continuously evolving. With the rapid expansion of digitalisation projects and advancements in laser scanning systems, the integration of photogrammetric imagery, high-resolution still images, renderings, and animations has become more prevalent.

Consequently, the need for CHO repositories has become increasingly clear. CH institutions can benefit significantly from such repositories, as they allow secure sharing of the scan data with administrators, clients, scholars, experts, and contractors worldwide; facilitating a range of needs in the field of cultural heritage such as scientific research and dissemination, documentation, conservation, site memory preservation, and restoration, public engagement and mediation, exhibitions, immersive experiences, films, etc.
This setup enables CH sites to lease computing resources, software, and systems on an as-needed basis via the Internet, thus reducing the expenses associated with traditional IT infrastructure. Additionally, there is a growing need for a cloud-based platform dedicated to the semantic enrichment and visual analysis of 3D models within the CH sector. Cloud infrastructures must evolve to become all-encompassing solutions that offer the necessary ICT support to host high-quality content and user-friendly graphical interfaces. From an infrastructure perspective, the primary artefacts of the 3D digitalisation process are the 3D model itself and the metadata and paradata~\cite{paradata-book} associated with it. Together, they constitute a CHO. Comprehensive and integrated management of all these components poses challenges for data management systems:

\begin{itemize}
\item archival and retrieval of resources that link various parts of a 3D object to its main structure,
\item tools for visual data analysis, data annotation, modification, validation, and importation/exportation of paradata and metadata,
\item advanced functionalities for importing and exporting 3D models in all available 2D and 3D formats, addressing related intellectual property rights (IPR) issues,
\item the capability to embed additional information into the 3D model structure, such as multimedia-linked data (audio, video, text, image),
\item single sign-on, federated authentication and authorisation mechanism to facilitate integration with research and cultural institutions,
\item distributed data management to allow federated management and contribution of infrastructure resources,
\item metadata-aware data management,
\item integrated, declarative data processing capabilities for creation of 3D models, processing pipelines, as well as metadata validation and processing,
\item sharing and publication of CHOs (particular models accompanied by metadata) as Open Data, or publishing to CH aggregators such as Europeana.
\end{itemize}

Such requirements call for designing a system that can handle the complexity of CHO objects and allow management and processing of both the data and the metadata, encompassing the whole process from the data ingestion to CHO publication.

A large number of infrastructures currently focus on 3D models, with recent reports~\cite{cieslik20203d}\cite{champion2020survey} summarising existing platforms, repositories~\cite{benardou2018cultural}\cite{nishanbaev2020web}\cite{nfdi}, online visualisation frameworks and formats~\cite{fernie20203d}. 
Digital infrastructures, services, and tools represent distinct yet interdependent components within the digital ecosystem. Digital tools are specialised software applications engineered for specific tasks, such as the generation of 3D models or the transformation of metadata. Conversely, a digital service constitutes a managed provision of a tool to external users, while digital infrastructures encompass one or more services and/or facilities that are required for the operation and delivery of digital capabilities. In~\cite{Münster2024}, authors distinguish four types of infrastructures for CHO and 3D models: data repositories, data aggregators, 3D data viewers, and virtual research environments. Repositories are collections of 3D models that are stored and made available for secondary use, the most widely used 3D repository globally being Sketchfab~\cite{sketchfab}. Data aggregators do not store 3D models directly, but instead compile databases to prepare combined data sets for data processing. For example, Europeana acts as a data aggregator by compiling its collection from data in national and regional libraries. 3D data viewers employ computer graphics to dynamically visualise and interact with 3D models, enabling functionalities such as rotation and zooming, e.g., Kompakkt~\cite{kompakkt}. Virtual research environments are web-based information systems that provide a comprehensive working environment for researchers, including various tools for analysis and comparison~\cite{candela2013virtual}. 

Currently, the 3D content is a relative newcomer to the field of cultural heritage~\cite{Huvila2025} and Europeana, which recently extended the EDM model to support 3D content~\cite{europeana-edm-3d}. The small number of existing 3D resources available in Europeana is made available by individual cultural heritage institutions~\cite{Tan_Raffin_2025}, mostly by the means of uploading the models to Sketchfab. At the same time, projects, e.g.~\cite{CoVHer}, aiming to facilitate standardization of digitalisation, processing, and publication of 3D cultural heritage either do not aspire to integrate with Eureopana or create complex and custom solutions~\cite{Quantin2025} that do not reach wider adoption. This shows a lack of infrastructure dedicated to hosting 3D resources, built with already existing tools and standards developed and supported by Europeana and the European Union.

Popularized by the EGI Foundation\footnote{https://www.egi.eu} within the scope of EU-funded projects,  Onedata~\cite{onedata-wrzeszcz-fgcs} is a global data management system, providing easy access to distributed storage resources and supporting use cases from personal data management to data-intensive scientific computations. It meets the most crucial of the above requirements related to the management of distributed data and metadata, data publication, integration with multiple identity providers, storage resource management, and integrated data processing. The system has already been used to provide distributed access and data processing of more than 4PB of archival cultural data sets from Polish national museums~\cite{onedata-orzechowski-fgcs}. It has also been used as a data aggregation and processing hub for data originating from different life-science experimental methods~\cite{onedata-czesi-life-siences}. 

Hence, in our solution, Onedata was chosen as a data and metadata management platform
for handling complex data and metadata of CHO via the introduction of automation workflows (AW) as an integrated part of the data management system.

\section{Cultural Heritage Objects Processing}
\label{section:CHO-solution}

The 3D digitalisation of CHO is an inherently complex multi-stage process~\cite{marinos-study}, and the final result consists of 3D models, metadata, and associated paradata. Moreover, different 3D digitised objects might require dedicated processing methods and procedures. This requires designing dedicated workflows for processing each type (or even each instance) of the 3D digitised object. 

\begin{figure}
\centering
\includegraphics[width=0.48\textwidth]{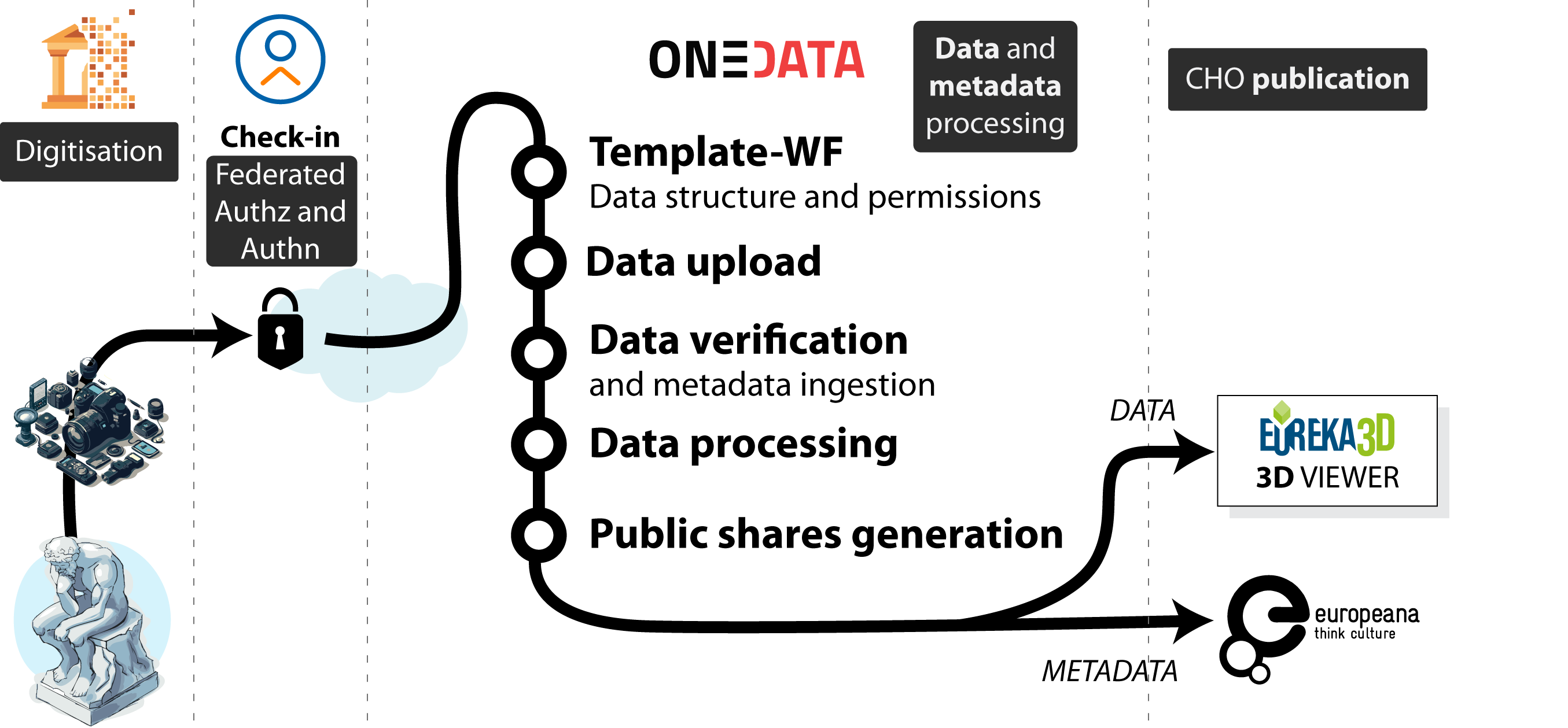}
\caption{Processing of CHO from digitization to publication. Each step is described subsequently in Section 3.}
\label{fig:automation-workflows}
\end{figure}

Although the responsibility for the process of performing 3D digitalisation rests solely on the entity of the content provider with access to the object and digitalisation equipment, the following digital processing can be standardised and documented within the data processing platform. This allows for the reproducibility of the 3D object processing and enables the reuse of common parts of the process for different 3D objects. Fig. \ref{fig:automation-workflows} shows the proposed generic workflow for a CHO processing starting with the digitalisation of the 3D model (\textit{Digitization}) followed by the authenticated (\textit{Federated Authz and Authn}) upload of the digitalisation results to the Onedata platform (\textit{Data and metadata processing}), processing of the results using dedicated automated workflows, and finally creating a public CHO record within Onedata, which can be accessed in 3D Viewer and embedded in Europeana (\textit{CHO Publication}). 

The following subsections describe each part of the generic workflow, excluding the digitalisation, in which results are treated as input for the generic workflow for a CHO processing.
\subsection{Federated Authentication and Authorisation}The underlying set of cloud resources used by the various CH institutions, including servers, storage, and services, should be protected against unauthorised access. This is critical to ensure the trust and integrity of the valuable digital assets managed in Cultural Heritage (mainly data, metadata, and paradata), and it is primarily achieved through the use of two mechanisms: authentication and authorisation. To fulfil this requirement, the EGI Check-in~\cite{egi-checkin} was chosen as its authentication and authorisation mechanisms are well integrated with identity management solutions used by the CHIs, allowing content providers to securely upload the results of the digitalisation process.

The EGI Check-in service provides identity and access management components that enable users to access community services and resources. It acts as an intermediate system connecting users, authentication servers, and services, offering users authenticated access to services and enabling single sign-on. It supports widely adopted standards and open technologies, including OIDC, OAuth, SAML, and X.509, which facilitates interoperability and integration with existing AAI services (responsible for managing the authentication and authorisation) of other Research Infrastructures and Research Communities. The EGI Check-in service has the following features:
\begin{itemize} 
\item provides increased productivity, with mechanisms such as single sign-on, and security,
\item accepts multiple federated authentication sources using different technologies,
\item is federated in eduGAIN\footnote{https://edugain.org} as a service provider, publishing REFEDS RnS\footnote{https://refeds.org} and Sirtfi\footnote{https://refeds.org/sirtfi, https://aarc-project.eu/policies/sirtfi} compliance,
\item provides a graphical user interface for user registration and management, which allows identity unification through account linking, as well as an API for programmatic user management,
\item can combine user attributes originating from various authoritative sources and deliver them to the connected service providers transparently. This process is conducted according to the GDPR,
\item is a member of the EOSC\footnote{https://eosc.eu} Access Federation.
\end{itemize}
In practical terms, EGI Check-in makes it possible to identify users who wish to access applications or data and to protect these resources from unauthorised access. This is done transparently to the user in a federated environment, which interconnects different cloud providers offering different resources in distinct geographic locations.
\subsection{Data and Metadata Processing}

The processing starts with a member of the content provider group from one of the CH institutions uploading the artefacts of the 3D digitalisation process: the 3D model and the associated metadata. However, before that happens, the content provider executes a dedicated CHO initialisation AW (\textit{Template-WF} workflow), creating a directory structure within the dedicated CH repository. The appropriate access policies are applied to the directory structure, allowing access only to the content provider. Now, the content provider can upload (\textit{Data upload}) the files containing the 3D model and the metadata to designated directories. Then, the AW for data and metadata validation (\textit{Data verification}) is executed. It validates the content of the 3D models and the files containing metadata. This step can be tailored to any 3D file format and metadata scheme, ensuring that uploaded data is correct. The metadata is then ingested into the Onedata platform, and the 3D files are annotated with it. This marks the logical creation of the CHO within the Onedata platform, allowing the CHO and its files to be searched in Onedata using the keywords in metadata, and later use this metadata to expedite the publication of the CHO. The 3D models that constitute the CHO are often large; this requires one to generate less complex (lower resolution textures, lower number of vertices) models (\textit{Data process}) of smaller size, suitable for later publication (\textit{Public shares generation}). This marks the creation of several derivative CHO objects, each containing the processed 3D model and the same metadata and paradata as the original CHO. 

In the last step, after verifying that the CHO object and its derivatives are correct, they can be published as single or separate public unlisted shares (only the person with the dedicated link can access them). Each share is annotated with the CHO metadata, and the additional paradata of the CHO processing workflow is added. This share can be ingested by the 3D Viewer, which extracts the 3D model or the list of its derivatives, allowing the end user to view it in a web browser.
\subsection{Cultural Heritage Object Publication}

Effective publishing of 3D digital CHOs requires an integrated solution that combines several complex functions: data ingestion, comprehensive management of data, metadata, and paradata, automated processing workflows, rich metadata annotation, persistent identifier (PID/DOI) minting, and public advertisement via protocols like OAI-PMH. In the realm of CH, these tasks must be executed in compliance with external standards to ensure interoperability with Europeana, the preferred platform for the CH communities in Europe.

\subsubsection{Europeana Data Model}

Europeana --- the central portal for digitised CH in Europe --- relies on the EDM, an RDF-based framework designed to aggregate metadata from diverse cultural institutions and promote the annotation of records with richer and more descriptive information. The EDM is the basis for building indices with aggregated metadata that fuel the search and discovery tools on the europeana.eu platform. Europeana does not store the digital objects themselves but retains direct references to the original data providers, ensuring that institutions maintain control over their digital assets while making them widely discoverable.
Europeana relies on OAI-PMH as its primary protocol for metadata harvesting, enabling structured and automated data collection from CHIs and aggregators. The use of EDM ensures interoperability across diverse cultural heritage domains, supporting metadata enrichment through \textit{Linked Open Data} principles. CH repositories willing to integrate with Europeana must satisfy the core technical requirements:

\begin{itemize}
    \item The 3D models must be published online and accessible in a viewer compatible with Europeana and embeddable using oEmbed.
    
    \item Metadata records must be available for harvesting via OAI-PMH in an XML format that is mapped to EDM, including information about both the cultural heritage object represented in the 3D model and the 3D model itself.
\end{itemize}

The early implementations of EDM primarily focused on textual and image-based resources. Now it is undergoing a refinement process, where the dedicated Europeana 3D Working Group\footnote{https://pro.europeana.eu/index.php/project/3d-content-in-europeana} is striving to accommodate 3D CHO better and improve metadata quality. A key area of development is paradata --- information about the digitisation process, methods, and post-processing changes. This is essential for assessing a 3D model’s suitability for different uses, such as conservation versus online visualization. 
Currently, the paradata lacks a means of structured description and is added as a reference in an EMD record to a dedicated paradata document. This allows for a more complete assessment of quality of the CHO as well as the digitization process itself.
The next revision of EDM introduces structured metadata and paradata fields to capture these details, ensuring better documentation of 3D digitization workflows. As the model evolves, compliance becomes more complex, requiring cultural heritage institutions to provide more detailed metadata and paradata.

The ongoing changes increase the complexity of compliance, making robust metadata and paradata management even more critical. %

\subsubsection{Onedata platform for integrated CHO management}

Onedata offers several desired capabilities, such as exposing public data collections with Dublin Core metadata, automatic PID/DOI minting, and record advertisement via OAI-PMH. However, they are insufficient for comprehensive CHO management. To fully satisfy Europeana’s EDM requirements and address the specific nuances of 3D digitised CHOs, several dedicated enhancements were introduced:\begin{itemize}
    \item To guide content providers and data stewards through the metadata curation process, a new graphical interface with a dynamic EDM form was developed (see Figure \ref{fig:europeana_data_model}). It relies on a pluggable architecture that ensures flexibility and ease of adjustment to the evolving EDM model. The UI automatically pre-fills the required EDM fields with ingested CHO metadata, which users can further enrich by linking additional paradata or supplementary data files. One of the machine-generated inputs is the \texttt{edm:isShownBy} field, which contains a URL to the Viewer application (oEmbed endpoint) with an inscribed reference to the published 3D model data, residing in Onedata. In this way, the interactive visualization of CHO is embedded in the Europeana portal. Whenever accessed, the 3D data, available publicly in read-only mode, is loaded into the Viewer application and rendered in the end-user's browser.    %

\begin{figure}[ht]
\centering
\includegraphics[width=0.5\textwidth]{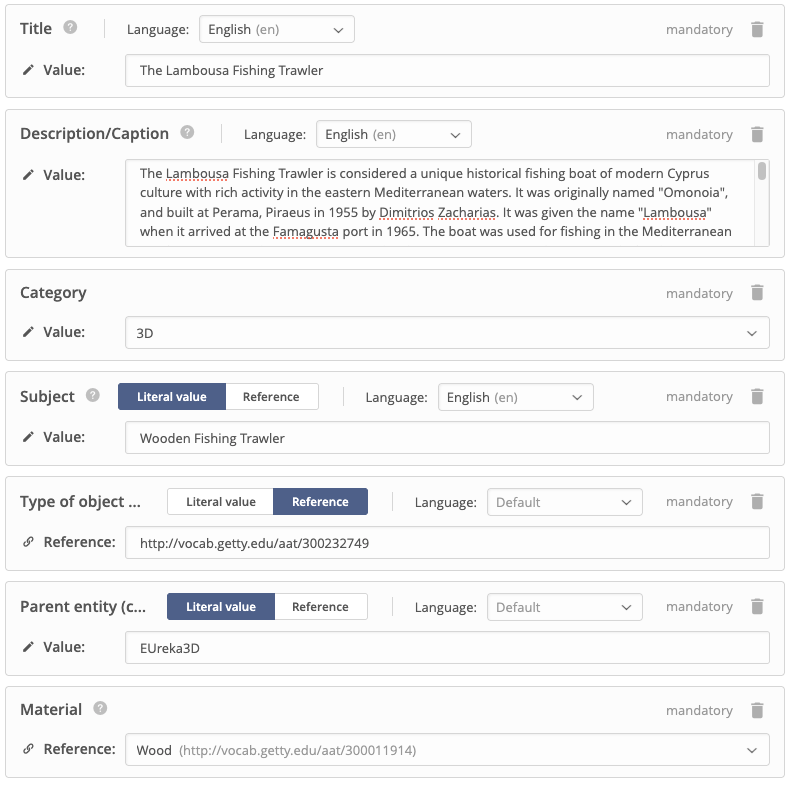}
\caption{A dynamic EDM editor that guides content providers and data stewards through the metadata curation process.}
\label{fig:europeana_data_model}
\end{figure}
    \item To ensure that the system can handle the automated annotation, processing, and publication based on EDM,
    the Onedata metadata engine was extended to support the new model across the system. Enhancements to the OAI-PMH service facilitate regular harvesting of CH records by Europeana, ensuring that CHOs remain discoverable and that possible metadata changes are reflected in the portal.

    \item To allow seamless and robust data publishing, repeatable and tedious tasks were automated and adapted for integration with middleware. Automatic assignment of persistent identifiers to CHOs benefits from enhanced integration with services such as EUDAT B2HANDLE\footnote{https://www.eudat.eu/service-catalogue/b2handle}. The REST API is used to automate the ingestion-to-publication process: in the case of the Bibracte\footnote{https://www.bibracte.fr/en} collection (about 500 3D models), a custom script was successfully employed to execute the entire process programmatically, reducing manual overhead and mitigating the risk of errors.
\end{itemize}\subsection{EGI DataHub}
\begin{figure}[ht]
\centering
\includegraphics[width=0.5\textwidth]{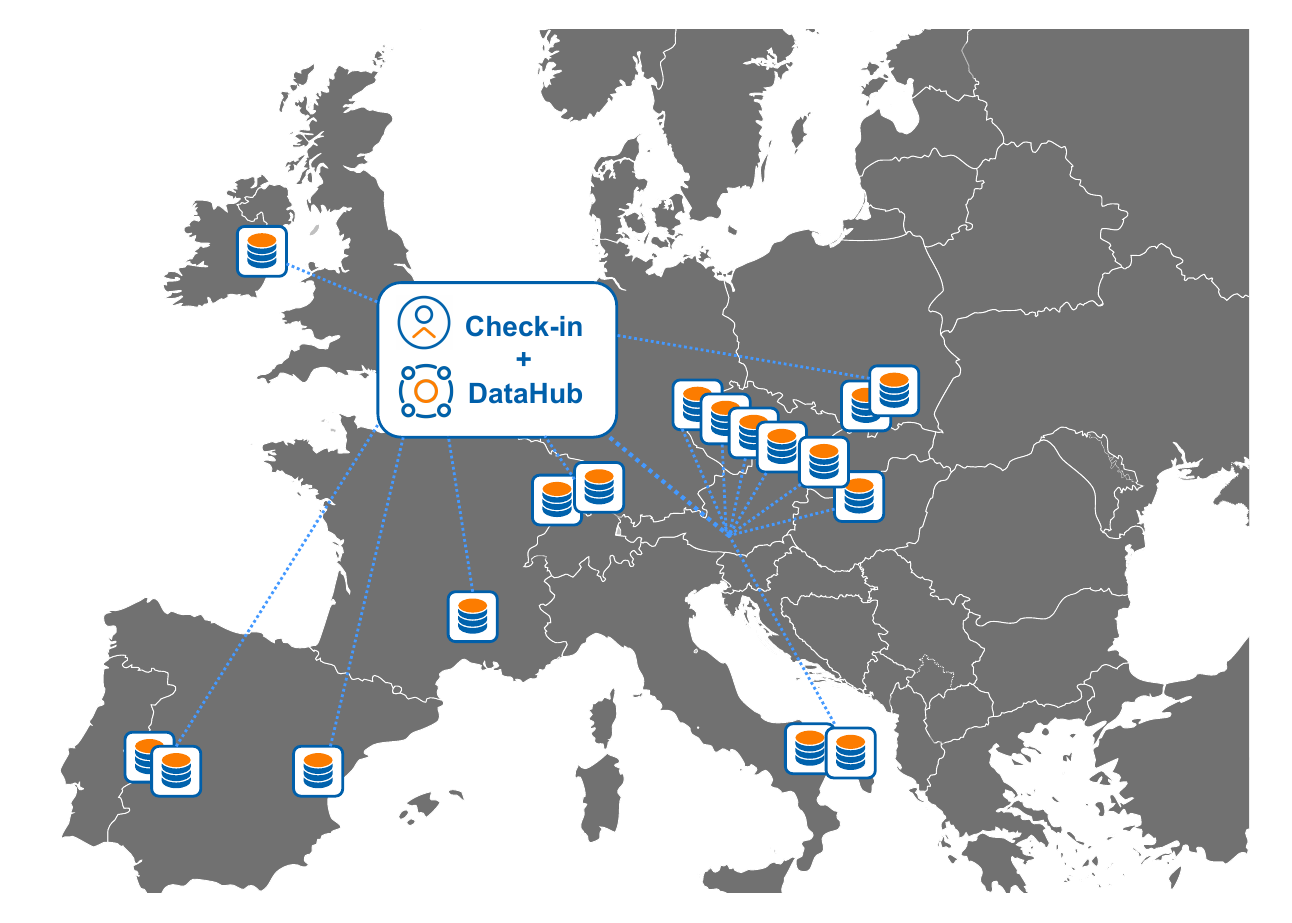}
\caption{EGI DataHub infrastructure.}
\label{fig:datahub-infrastructure}
\end{figure}

Validation of the above enhancements was carried out within the EGI DataHub ecosystem --- a Europe-wide platform for distributed data management, based on the Onedata platform. It is one of the EGI core services\footnote{https://www.egi.eu/service/datahub/}, establishing a distributed network of data providers (cf. Figure \ref{fig:datahub-infrastructure}). As of February 2025, it brings together 17 data centres, jointly hosting over 2200 distributed data collections (totalling ~2.3~PB of allocated storage space) and catering to many scientific projects. It has been chosen due to its tight integration with EGI Check-in and multi-national scale. Thanks to the EGI's engagement in CH-related EU projects, it was possible to align the efforts with the CH community and specialists. 

The deployment and evaluation on EGI DataHub confirmed that the extended Onedata platform provides a comprehensive, integrated solution for CH data management and publication. Users can perform tasks ranging from ingestion and organisation of the data, through processing and metadata curation, and finally to publication. The CHOs are annotated with rich EDM metadata, automatically assigned a PID thanks to DataHub's integration with the EUDAT B2HANDLE service, advertised in OAI-PMH, and periodically harvested by the Europeana Metis ingestion tool. The integration of the proposed solutions in the EGI Datahub ecosystem has served as a successful proof-of-concept, accommodating many CHOs that are now publicly available and discoverable on europeana.eu, together with their 3D visualisations. Furthermore, the wide European network of EGI DataHub allows for further denomination of CHOs within the EGI DataHub ecosystem, and thanks to EGI Check-in the system is easily accessible to a large number of European CHIs.
\section{Automation workflows with Onedata}
\label{section:onedata}
Managing data and metadata in distributed environments often involves repetitive, time-consuming, and error-prone tasks when performed manually, making scalability and consistency challenging. This is particularly problematic in domains like CH, where users may lack technical expertise but require pipelines for data curation, transformation, and analysis. There is a need for tools that can ensure reproducible and unified data processing results, support continuous refinement of procedures, and empower non-technical users to perform complex data management tasks. Currently, there is a wide choice of both commercial and open-source workflow systems that can automate data processing and management tasks, each offering various features and capabilities. However, using such systems alongside a separate data management platform is problematic. Non-technical users may find it overwhelming to navigate multiple tools to achieve their objectives, and differences in standards and data formats between platforms can create barriers when transferring data. These issues make it difficult to achieve a seamless working environment and highlight the need for a more consolidated approach to automation.

Integrated automation workflows within the Onedata system address these challenges by offering a seamless and efficient means to automate tasks directly within the platform. Users can compose and execute sequences of actions on selected data collections without relying on external tools. Onedata orchestrates the execution of these workflows, while users retain full control over the logic of individual data processing functions, known as lambdas, which define how data is manipulated, transformed, or analysed.

The term \textit{integrated} signifies that these workflows are executed within Onedata's built-in automation engine. This tight integration allows lambdas to directly access and operate on user data, with data location and distribution abstracted by the Onedata layer. The developer of a lambda function has access to a wide range of Onedata APIs, including data transfers, Quality of Service rules, custom metadata management, public sharing, and archiving, ensuring comprehensive data management capabilities. Additionally, workflows are executed in a containerised fashion on a Kubernetes cluster, facilitating on-demand and scalable execution of tasks. The execution layer is a generic concept --- while the current implementation uses OpenFaaS~\cite{openfaas} to run the lambda containers, other task execution frameworks can be plugged in. 

\begin{figure}[ht]
\centering
\includegraphics[width=0.5\textwidth]{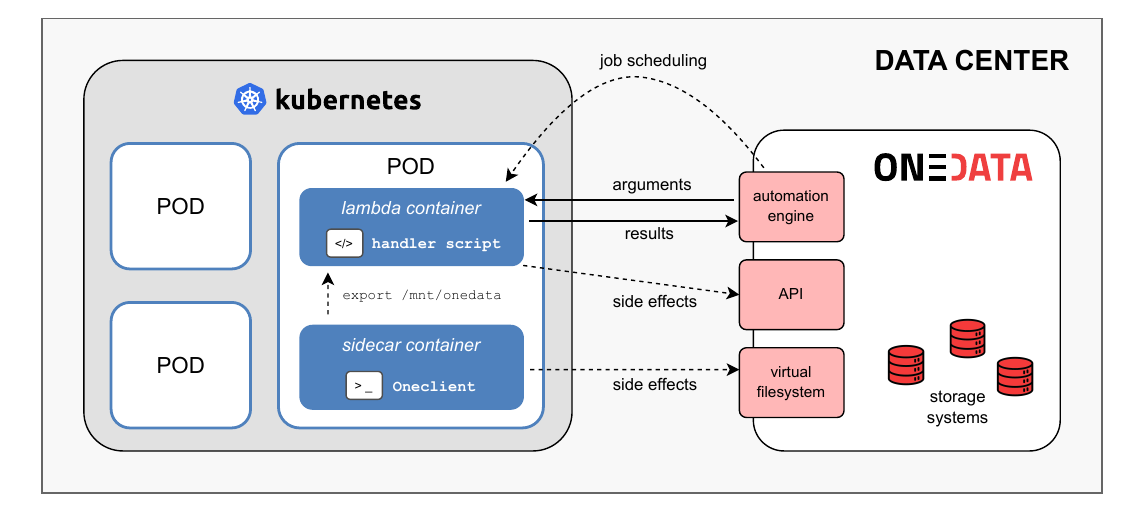}
\caption{Architecture of Onedata's automation engine in the scope of a data center.}
\label{fig:workflow-engine-architecture}
\end{figure}
The seamless integration of the lambda execution environment with the Onedata filesystem is achieved through the use of sidecar injection within Kubernetes pods (cf. Figure \ref{fig:workflow-engine-architecture}). Each lambda runs as a container within a pod, alongside a sidecar container that hosts the Oneclient application (Onedata client based on the fuse\cite{fuse} library). The sidecar mounts the Onedata filesystem onto the pod's local filesystem, creating a shared mount point accessible to both containers. This design allows the lambda to interact with data in Onedata natively, as if it were operating on a local POSIX filesystem. To support scalability, multiple instances of these pods are deployed on demand to match the workload.

\begin{figure}[ht]
\centering
\includegraphics[width=0.45\textwidth]{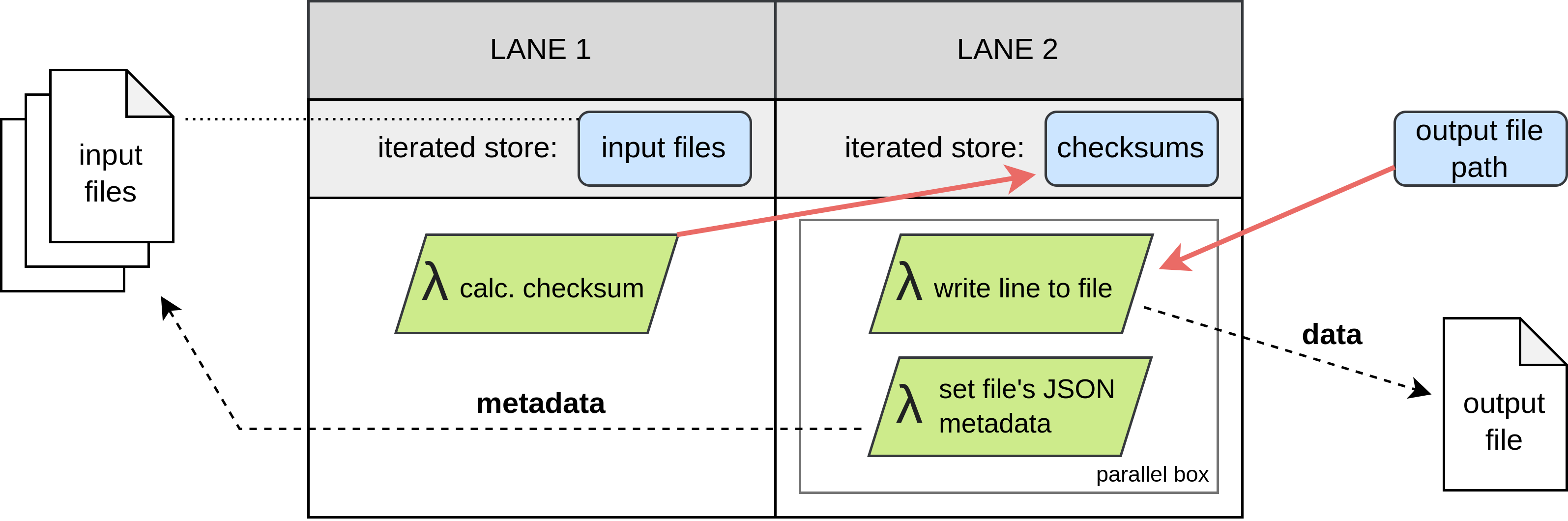}
\caption{Workflow schema example, showing its basic anatomy.}
\label{fig:workflow-schema-anatomy}
\end{figure}

Automation workflows in Onedata are based on three main concepts: lambdas, workflow schemas, and inventories. A lambda constitutes a task with arbitrary logic that takes input arguments and runs a custom-implemented handler that produces output results. The specification of arguments and results is included in the lambda definition, together with their data types. During execution, inputs and outputs are encoded to JSON during transmission to and from the handler, making it easy to implement in any programming language. Many lambda functions are designed to have side effects by interacting with the Onedata filesystem or APIs. For example, a lambda might take two input arguments --- \texttt{output file path} and the list of \texttt{lines to write} --- to create a new file at the specified path, write the requested lines to it, and return the new file's ID as the output (see Figure \ref{fig:workflow-schema-anatomy}).

Workflow schemas can be perceived as blueprints for data processing pipelines, as depicted in Figure \ref{fig:workflow-schema-anatomy}. They run a series of lambdas on a set of inputs to produce outputs and side effects. Intermediate products of processing steps can be used as inputs for consecutive tasks --- to that end, the so-called stores are used. They hold collections of typed data and act as sinks for the results produced by lambdas. Stores can also be used to hold the user input required for starting an execution or the final results of the whole run. A workflow schema consists of one or more lanes that are processed sequentially. 
Each lane has one or more lambdas and one designated store it iterates over, running all the lambdas for each element in the store. Within a lane, the lambdas are executed sequentially, unless they are placed in a parallel box, allowing them to be run in parallel.

An inventory is a registry of lambdas and workflow schemas that can be made accessible to multiple users and user groups, allowing them to share and reuse common definitions. The privileges on the inventory level regulate who can create, modify, or use the definitions. They should be set to reflect the roles of users in a certain group or project. For instance, the creation of lambdas and workflow schemas requires programming skills and technical expertise, and they should not be tampered with by non-technical users. They, however, constitute the audience that will be executing the workflows and should communicate with the implementers to constantly refine the definitions. Optimally, a workflow schema should hide the complexity of internal tasks and provide a clear abstraction to the end-users so they understand its purpose and outcomes.

The example in Figure \ref{fig:workflow-schema-anatomy} shows a workflow schema that takes a set of input files and calculates their checksums, storing them as JSON metadata of corresponding input files and writing a summary to an output file. The stores \textit{input files} and \textit{output file path} hold the initial user input required to run the workflow. The \textit{checksums} store serves both as an intermediary container for data and a final artefact of the execution, storing \texttt{\{file, checksum\}} pairs. In the second lane, this store is iterated over, and for every pair, a new line to the summary output file is written, and a JSON metadata entry with the checksum is assigned to the input file.
\section{Integration and deployment in EUreka3D Project}

\begin{figure*}
\centering
\includegraphics[width=0.7\textwidth]{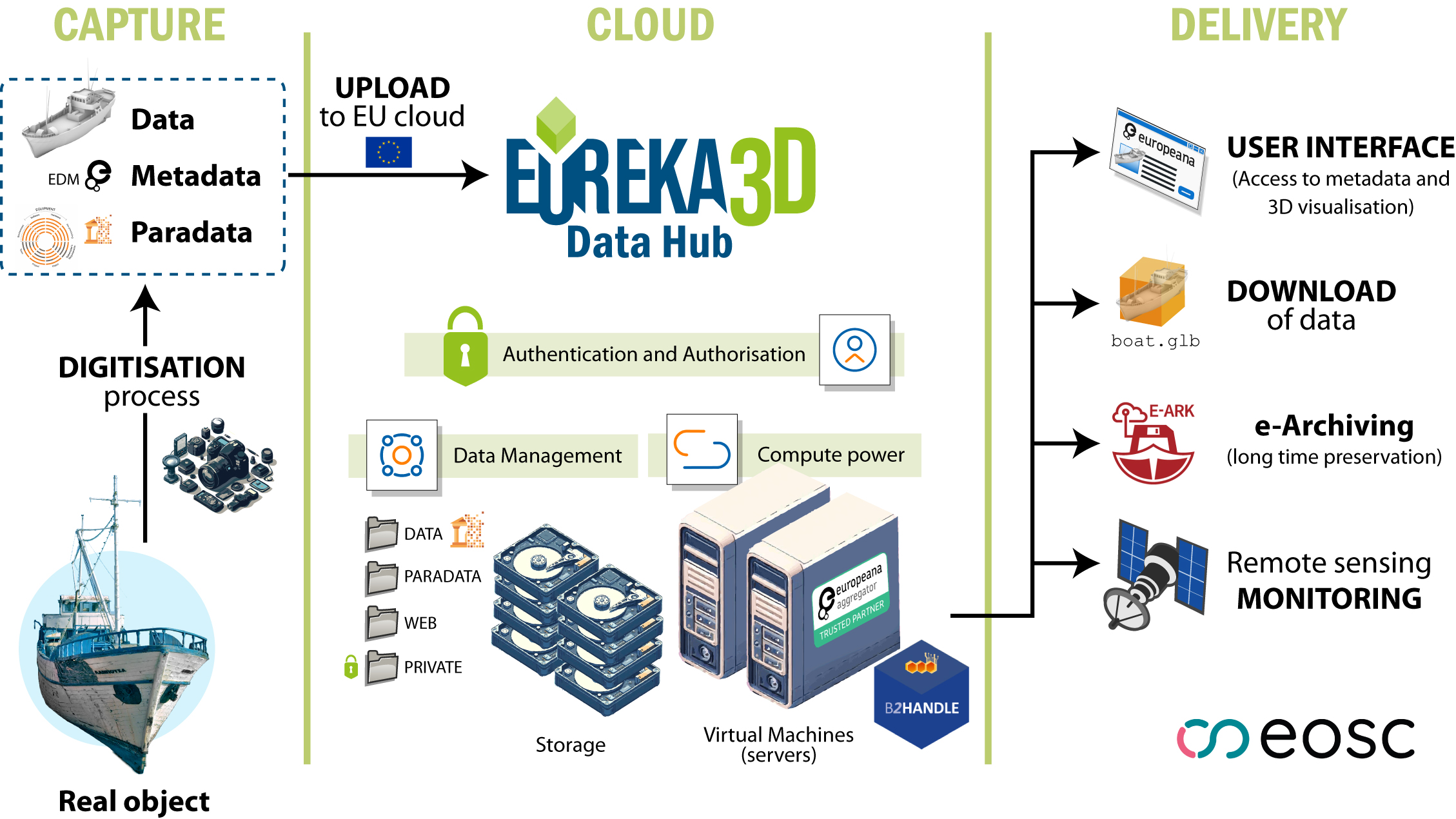}
\caption{CHO capture, processing, and delivery workflow shown as part of EUreka3D.}

\label{fig:eureka3d}
\end{figure*}The EUreka3D \cite{eureka3d2025} project addresses the growing need for enabling the 3D digital transformation of the Cultural
Heritage sector. Museums, galleries, libraries, archives, and archaeological sites are witnessing a context
of technological change and need to review and modernise their internal processes, from digital capture to
end-user access and re-use. This context of change was fuelled by the EU Recommendation
to digitise in 3D by 2030 all monuments and sites deemed at risk and half of the most physically
visited cultural and heritage monuments, buildings, and sites. The recommendation sets ambitious targets
on Member States and the CHIs in Europe, but starts from a base where standards and knowledge on 3D
digitization and coherent methodologies, infrastructures, and processes are lacking. 

To achieve its goals, EUreka3D brought together content providers who were at different stages of their 3D transformation journey, from first steps with objects in controlled environments to highly experienced digitisation experts with complex
objects in uncontrolled environments. 

The EUreka3D developed a pilot action to experiment with innovation in the workflow of CHIs, especially to accommodate the needs and requirements connected to managing and sharing 3D collections. In the EUreka3D  partners undertook 3D digitisation of a diverse range of objects, from museum artefacts to archaeological sites, all following the technical specifications based on the relevant Study on Quality in 3D Digitisation of Tangible Cultural Heritage~\cite{marinos-study}. The 3D models were accompanied by relevant metadata and paradata to provide comprehensive documentation of the objects represented in 3D. 

The complex process covered in the EUreka3D project is graphically illustrated in Figure \ref{fig:eureka3d}, which consists of three main phases:

\begin{itemize}
\item Capture: the actual digitisation process for the cultural objects and the creation of data, metadata, and paradata.
\item Cloud Infrastructure: where produced files (data, metadata, and paradata) are uploaded on the cloud
and safely accessible with different levels of authorisation for access, with open access policy
preferred.
\item Delivery: the models are visualised in a viewer compatible with Europeana, the metadata are
inputted and converted in the EDM, and the paradata are linked as open access
files. At the end of this phase, aggregation to Europeana and publication in europeana.eu website
eventually happens.
\end{itemize}
The implementation of the cloud and delivery phases is based on: workflow automation, integrated data, metadata, and paradata management, federated authentication and authorization, the CHO publication process to Europeana, and a distributed Data Hub infrastructure. The work presented in this article was heavily influenced by the EUreka3D project requirements and was used by cultural content providers to accomplish the project goals. 
\section{Evaluation}
\label{section:evaluation}

The described AW for the CH domain was developed as part of the EUreka3D project. For their evaluation, we chose three distinct 3D models digitised as part of EUreka3D by the Cyprus University of Technology: the small movable CHO of a \textit{figurine}, the medium-sized immovable CHO of the \textit{Lambousa Fishing Trawler}, and the large complex immovable CHO of the \textit{Church of the Holy Cross}. In this section, we discuss the importance of model preparation, briefly describe the historical background, the digitisation processes, provide the detailed EDM records for each object, and discuss model processing using AW.  

\begin{table*}[htbp]
\centering
\begin{adjustbox}{width=\textwidth}
\begin{tabular}{|l|c|c|c|c|c|c|}
\hline
 & \textbf{Geometrical Survey - Raw Data} & \multicolumn{2}{|c|}{\textbf{Geometrical Survey - Post-processed data}} & \multicolumn{3}{|c|}{\textbf{3D CAD Model - Raw Data}} \\
\hline \hline
\textbf{Type} & Photogrammetry & Terrestrial Laser Scanning  & Downsampled and Aligned  & Raw Data & Post-processed data  & Workflow processing \\
 & & (TLS) & Photogrammetry with TLS &  &  for Web &  \\
\hline
\textbf{Hardware} & DJI Mini 3 pro  & Z+F Imager 5016 & PC Hardware & PC Hardware & PC Hardware & Onedata \\
 & and Sony A7 IV Camera & &  &  &  &  \\
\hline
\textbf{Software} & Reality Capture 1.2.2 & Z+F Laser Control Scout & CloudCompare & Rhinoceros & Blender & Blender \\
\hline
\textbf{Overall Time} & 8 working days & 1 working day & 2 working days & 115 working days & 1 working day & 3h \\
\hline
\textbf{Software Processing time} & 2 h and 10 min & 1 hr & 1.5 hr & 57 h & 15 min & 15-9 min\\
\hline
\textbf{No. of points} & 26,039,030 & 54,495,372 & 19,164,749 & & & \\
\hline
\textbf{No. of polygons} & 30,001,378 & & & 10,078,258 & 1,365,772 & 1,365,772-195,110\\
\hline
\textbf{Formats} & OBJ & E57 & E57 & OBJ & GLB & GLB \\
\hline
\textbf{File Size} & 6,639,738 KB & 1,657,171 KB & 563,169 KB & 1,497,417 KB & 162,107 KB & 162,107-23,158 KB \\
\hline
\textbf{Automation Level} &  \multicolumn{5}{|c|}{Manual} & Automatic \\

\hline
\hline
\textbf{Europeana Link} & \multicolumn{6}{c|}{https://www.europeana.eu/en/item/1268/21\_15123\_DKf8oUnT} \\
\hline
\end{tabular}
\end{adjustbox}
\caption{The Lambousa Fishing Trawler - 3D Data Statistics.}
\label{tab:lambousa_data}
\end{table*}

\subsection{Models Preperation}

The acquisition of digital data for tangible CH assets presents varying degrees of complexity, contingent upon the specific user scenario and environmental context.  Simple scenarios may involve small, readily accessible artifacts in controlled environments, such as museums or laboratories. Conversely, complex scenarios encompass challenging contexts such as underwater shipwrecks or objects located within caves, where accessibility and environmental factors significantly increase the difficulty of data acquisition.  
High-fidelity digitization, encompassing both geometric and photometric data, along with comprehensive metadata and paradata (contextual information related to the object and the digitization process), is paramount. This approach maximizes the potential for long-term preservation, facilitates diverse applications (research, education, conservation, etc.), and enhances accessibility for a wider audience.
Digitization projects require interdisciplinary teams of experts utilizing specialized high-resolution capture systems in challenging and dynamic environments, yielding comprehensive and versatile 3D models with enhanced longevity and broader applications across diverse stakeholder communities.
The models presented in this section have undergone the digitisation process in line with those standards. Tables \ref{tab:lambousa_data}, \ref{tab:shm_data}, and \ref{tab:church_data}  present detailed information on the different aspects on complexity of objects (number of points, polygons, file size) and the digitisation process (photogrammetry, terrestrial laser scanning, downsampling etc.). Furthermore, there are two distinct levels of automation:
\begin{itemize}
    \item manual, where a team of experts from CH domain proceeds with each step digitisation process ensuring highest quality standards, which is reflected by a significantly long \textit{Overall Time} of \textit{Raw Data} processing and post-pro\-ces\-sing of point cloud data resulting in a first 3D model of an object,
    \item automatic, where experts from the computer science domain automate the post-processing phase of a model, allowing for the generation of models of various quality and size.
\end{itemize}
The last row of the tables contains the web link to the Europeana record for each object, which contains a working 3D Model presented using an EUreka3D viewer and detailed information on metadata and paradata.

\subsection{Lambousa Fishing Trawler}

\begin{table*}[htbp]
\centering
\begin{adjustbox}{width=\textwidth}
\begin{tabular}{|l|c|c|c|}
\hline
 & \textbf{Geometrical Survey - Raw Data} & \multicolumn{2}{|c|}{\textbf{3D CAD Model}} \\
\hline \hline
\textbf{Type} & Photogrammetry & Post-processed data for Web & Workflow processing \\
\hline
\textbf{Hardware} & Sony A7iii camera, LED lighting Yongnuo Digital YN600/300 Air and a Lightbox & PC Hardware & Onedata \\
\hline
\textbf{Software} & Agisoft Metashape & Blender & Blender \\
\hline
\textbf{Overall Time} & 4 h & 4 h & 2 h \\
\hline
\textbf{Software Processing time} & 42 min & 28 s & 20-15 s \\
\hline
\textbf{No. of points} & 3,330,971 & 746,814 & 746,814-106,687 \\
\hline
\textbf{No. of polygons} & 248,938 & 248,938  & 248,938-35,562 \\
\hline
\textbf{Formats} & PSX & OBJ & OBJ \\
\hline
\textbf{File Size} & 1,230,000 KB & 41,534 KB & 41,534-5.933 KB \\
\hline
\textbf{Automation Level} &  \multicolumn{2}{|c|}{Manual} & Automatic \\
\hline
\hline
\textbf{Europeana Link} & \multicolumn{3}{c|}{https://www.europeana.eu/en/item/1268/21\_15123\_gPUY00xiD} \\
\hline
\end{tabular}
\end{adjustbox}
\caption{Sculpture, Figurine - Cypriot Artefacts Collection from the Museum of Mediterranean and Near Eastern Antiquities in Stockholm, Sweden - 3D Data Statistics.}
\label{tab:shm_data}
\end{table*}

\begin{table*}[htbp]
\centering
\begin{adjustbox}{width=\textwidth}
\begin{tabular}{|l|c|c|c|c|c|c|}
\hline
 & \textbf{Geometrical Survey } & \multicolumn{2}{|c|}{\textbf{Geometrical Survey - Post-processed data}} & \multicolumn{3}{c|}{\textbf{3D CAD Model}} \\
 & \textbf{ - Raw Data} & \multicolumn{2}{|c|}{\textbf{}} & \multicolumn{3}{c|}{} \\
 \hline \hline
\textbf{Type} & Photogrammetry & Terrestrial Laser Scanning  & Aligned Photogrammetry  & Raw Data & Post-processed data & Workflow processing \\
&  & (TLS) &  with TLS & & for Web &  \\
\hline
\textbf{Hardware} & DJI Phantom 4 pro  & Z+F IMAGER 5016 & & PC Hardware & PC Hardware & Onedata \\
 &  and Sony A7iii Camera &  & & &  & \\
\hline
\textbf{Software} & Reality Capture 1.4.2 & Z+F Laser Control Software & CloudCompare  & Autodesk Revit & Autodesk Revit & Blender \\
 &  &  & and Autodesk Recap & &  &  \\
\hline
\textbf{Overall Time} & 3 working days & 2 working days & 3 working days & 43 working days & 2 min & 2h \\
\hline
\textbf{Software Processing time} & 3 h and 25 min & 5 h & 4 h & 10 h & 14 s & 14-10 s \\
\hline
\textbf{No. of points} & 33,962,883 & 88,290,000 & 122,252,883 & N/A (BIM Geometry) & & \\
\hline
\textbf{No. of polygons} & 32,820,819 & & & N/A (BIM Geometry) & 468,747 & 468,747-66,963 \\
\hline
\textbf{Formats} & OBJ & E57 & RCP & RVT & OBJ & OBJ \\
\hline
\textbf{File Size} & 1,672,228 KB & 2,269,128 KB & 3,040,000 KB & 1,048,164 KB & 74,343 KB & \\
\hline
\textbf{Automation Level} &  \multicolumn{5}{|c|}{Manual} & Automatic \\
\hline
\hline
\textbf{Europeana Link} & \multicolumn{6}{c|}{https://www.europeana.eu/en/item/1268/21\_15123\_Jz6REeQV} \\
\hline
\end{tabular}
\end{adjustbox}
\caption{The Church of the Holy Cross - 3D Data Statistics.}
\label{tab:church_data}
\end{table*}

\subsubsection{Historical Background}
Lambousa is a wooden vessel of the liberty style, characterized by its ellipsoid stern, developed after World War II~\cite{damianidis1998vernacular}. It was constructed in 1955 in Greece and registered in Cyprus in 1965. Measuring 25 meters in length, it is equipped with a diesel KELVIN engine and two masts, initially intended for sailing but later exclusively utilized for fishing. It stands as a distinctive heritage monument, being the last remaining trawler of its kind that still exists throughout the entire Panhellenic region. 
The vessel was dry-docked in 2020 for maintenance, financed by a European grant aimed at its conservation~\cite{nicolaou2025restoring}.

\subsubsection{Digitisation Rationale}
During the restoration process, a comprehensive digital documentation initiative was undertaken to protect its history and to facilitate the utilisation of the data by the multi-disciplinary society. Specifically, the three-dimensional model is accessible in multiple formats, including 3DM, OBJ, STL, E57, and GLB, thereby serving various sectors such as Naval Engineering, Archaeology, 3D Fabrication, Web Visualization, and Gaming. The paradata of the digitisation was recorded according to the VIGIE 2020/654 guidelines, while the metadata is in alignment with EDM, thus enabling effective aggregation within Europeana. To enhance the documentation of knowledge to its fullest potential, an online platform was also established, which features an e-story book, narratives from archaeologists and marine engineers engaged in the restoration, along with interactive educational games.

\subsubsection{Digitisation Process}
\begin{figure}
\centering
\includegraphics[width=0.4\textwidth]{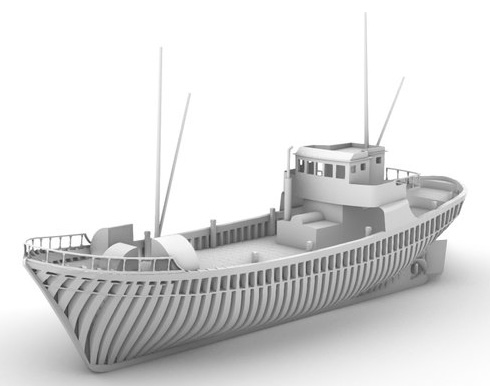}
\caption{3D model of the \textit{Lambousa Fishing Trawler}.}
\label{fig:lambousa-model}
\end{figure}

A holistic digitisation was performed using the Memory Twin methodology, which is an innovative extension of the Digital Twin concept, especially in DCH. Unlike Digital Twins, Memory Twins include a wider range of information, such as paradata, metadata, and data. Paradata adds depth by documenting the rationale and processes behind data creation, offering insights into the digitisation journey. Metadata enhances digital content's significance by clarifying the information. Data consists of digital representations of cultural artefacts, encompassing their holistic knowledge, 3D models, images, and documents that depict their physical traits. Moreover, the Memory Twin approach allows the incorporation of intangible data, like maritime craftsmanship techniques, into tangible forms such as 3D models~\cite{ioannides2021digital}~\cite{huvila2025imperative}~\cite{ioannides2025integrating}. 
The holistic digitisation was part of H2020 ERA Chair MNEMOSYNE~\cite{mnemosyne2025} and Digital Europe EUreka3D projects which aimed to preserve the vessel's history. This process commenced in January 2023 with a UAV photogrammetric survey using a DJI Mini 3 Pro\cite{dji2025mini3pro} and Reality Capture software~\cite{realitycapture2025} to process 1100 images. This resulted in a georeferenced mesh model with 27,363,867 faces. CloudCompare~\cite{cloudcompare2023} was used to refine the photogrammetric mesh data, downsampling it to 5,000,346 points. A subsequent survey in October 2023 utilized TLS with the Z+F Imager 5016 Laser Scanner~\cite{zoller2025imager} to capture the geometry of wooden structural elements, resulting in 14,164,403 points. The TLS point cloud was aligned with the photogrammetric data for comprehensive processing. This data was processed using AutoCAD~\cite{autocad2025} and Rhinoceros~\cite{rhinoceros2022} CAD software to generate shipbuilding lines and a 3D model. Vertical cloud sections were created, and splines were developed to form a network of curves for the hull. The hull surface was then developed, allowing for the creation of solid geometry. Using this geometry, the deck, frames, deck beams, cap rail, keel, and stern were shaped. Additional features like the fish hold, cabin, and masts were created by tracing curves over the point cloud. The result is a closed 3D CAD model (see Fig.~\ref{fig:lambousa-model}) with NURBS geometry at Level of Detail 400. A deviation analysis evaluated the NURBS geometry's accuracy against the point cloud, showing deviation between -10 to 12mm~\cite{ioannides2024d21}.
Furthermore, the intangible data was acquired through comprehensive research into archival materials, alongside interviews with maritime specialists. Specifically, the craftsmanship and maritime customs were recorded through a collection of narratives from the archaeologists, craftsmen, and engineers engaged in the project, in addition to historical documentation and photographic evidence.

\subsection{The Cyprus Collection at Medelhavsmuseet}

\subsubsection{Historical Background}
The Cyprus Collection at Medelhavsmuseet (the Museum of Mediterranean and Near Eastern Antiquities) comprises approximately 1,500 archaeological discoveries covering a timeframe that extends beyond 7,000 years from the Stone Age through to Imperial Rome. These discoveries demonstrate the occurrence of frequent and prolonged exchanges between Cyprus and other cultures throughout the eastern Mediterranean region. The majority of the artefacts displayed in this exhibition are derived from the Swedish Cyprus Expedition conducted between 1927 and 1931, during which excavations took place at more than twenty sites across the island. 

\subsubsection{Digitisation Rationale}
The Swedish Cyprus Expedition Digitisation Initiative started in January 2024 within the framework of a significant bilateral agreement between the Cyprus University of Technology and Medelhavsmuseet. This agreement aims to digitise the largest collection of Cypriot antiquities outside of Cyprus.
The digitisation of this collection holds the promise of unique research opportunities, establishing new educational material and reuniting these artefacts, bridging physical distances and serving as an inspiring example of exceptional cooperation in the domain of cultural diplomacy.

\subsubsection{Digitisation Process}
\begin{figure}
\centering
\includegraphics[width=0.15\textwidth]{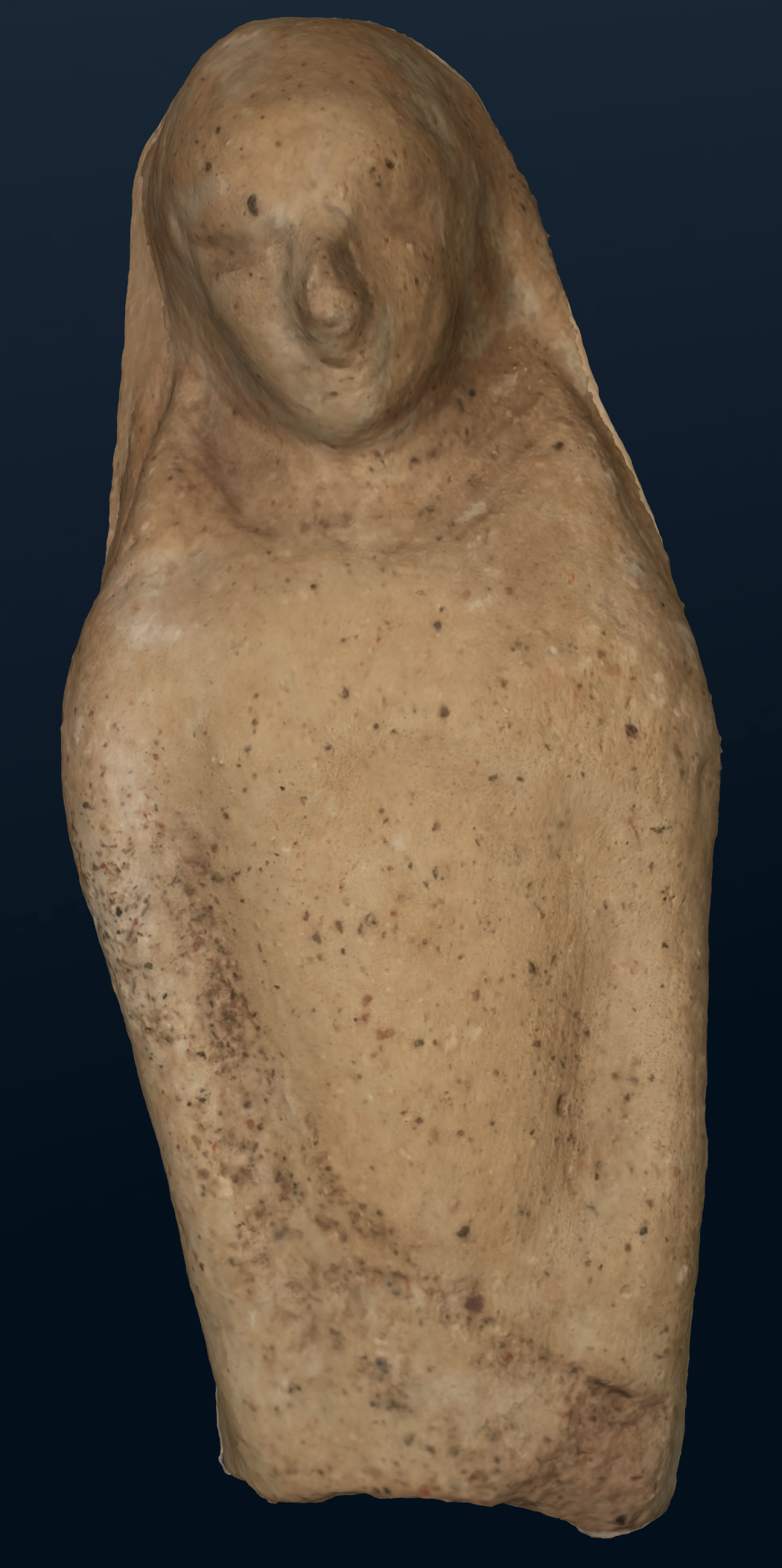}
\caption{3D model of the \textit{Sculpture, Figurine}.}
\label{fig:figurine}
\end{figure}

The digitisation which includes paradata, metadata, and 2D/\-3D data, was based on the VIGIE 2020/654 Study. It began on 14/01/2024 until 20/01/2024. For the small artefacts, photogrammetry was used and for bigger artefacts that cannot be moved, handheld scanning was used. A dedicated room in the museum was used for the digitisation. The artefacts had to be moved from the museum staff to the room. The equipment used for the photogrammetry was the Sony A7iii camera, two diffused LED lighting Yongnuo Digital YN600 Air with tripods and one YN300 Air adjusted to the camera, a lightbox, and a turntable. The software used for processing the images was Agisoft Metashape. Through this expedition, 31 artefacts have been digitised. Twenty artefacts were aggregated through the National Aggregator of Cyprus, and two (3D model of one of the Sculptures, see Fig.~\ref{fig:figurine}) from the EUreka3D infrastructure.

\subsection{The Church of the Holy Cross}

\begin{table*}[htbp]
    \centering %

    {%
        \setlength{\extrarowheight}{2pt} %
        \arrayrulecolor{ruleGray}        %
        \setlength{\arrayrulewidth}{0.5pt} %

        \scriptsize %
        \begin{tabularx}{\linewidth}{@{} P{3.5cm} L @{}} %

        \multicolumn{2}{@{} >{\columncolor{headerGray}} p{\dimexpr\linewidth-2\tabcolsep} @{}}{%
          \parbox[t]{\linewidth}{\vspace{0pt}\textbf{Cultural Heritage Object}\vspace{2pt}\par This section contains information about the physical Cultural Heritage Object.\vspace{4pt}}%
        } \\
        \hhline{>{\arrayrulecolor{ruleGray}}--}
        Title & The Lambousa Fishing Trawler \\
        \hhline{>{\arrayrulecolor{ruleGray}}--}
        Description & The Lambousa Fishing Trawler is considered a unique historical fishing boat of modern Cyprus culture with rich activity in the eastern Mediterranean waters. It was originally named "Omonoia", and built at Perama, Piraeus in 1955 by Dimitrios Zacharias. It was given the name "Lambousa" when it arrived at the Famagusta port in 1965. \\
        \hhline{>{\arrayrulecolor{ruleGray}}--}
        Category & 3D \\
        \hhline{>{\arrayrulecolor{ruleGray}}--}
        Subject & Wooden Fishing Trawler \\
        \hhline{>{\arrayrulecolor{ruleGray}}--}
        Type of object & \url{https://vocab.getty.edu/aat/300232749} \\ 
        \hhline{>{\arrayrulecolor{ruleGray}}--}
        Creator of the original object & Dimitrios Zacharias \\
        \hhline{>{\arrayrulecolor{ruleGray}}--}
        Creation date of the original object & 1955 AD \\
        \hhline{>{\arrayrulecolor{ruleGray}}--}
        Dimensions with units & Length Over All (L.O.A): 24,95m Maximum Breath (M.B.): 6,54m Length of the Keel (L.K.): 19,93m Length of Bow (L.B.): 3,73m Length of Stern (L.S.): 2,72m Depth of Bow (D.B.): 4,91m Depth of Stern (D.S.): 5,31m Height of the Mid-frame (M.D.): 4,02m\\
        \hhline{>{\arrayrulecolor{ruleGray}}--}
        Parent entity (collection, object, site) & E-RIHS 3D \\
        \hhline{>{\arrayrulecolor{ruleGray}}--}
        Material & \url{https://vocab.getty.edu/aat/300011914} \\ 
        \hhline{>{\arrayrulecolor{ruleGray}}--}
        Material & \url{https://vocab.getty.edu/aat/300010900} \\ 
        \hhline{>{\arrayrulecolor{ruleGray}}--}

        \multicolumn{2}{c}{} \\[-1.5ex] %

        \multicolumn{2}{@{} >{\columncolor{headerGray}} p{\dimexpr\linewidth-2\tabcolsep} @{}}{%
          \parbox[t]{\linewidth}{\vspace{0pt}\textbf{Digital Object}\vspace{2pt}\par This section contains information about the digital representation of the Cultural Heritage object (e.g. the 3D model).\vspace{4pt}}%
        } \\
        \hhline{>{\arrayrulecolor{ruleGray}}--}
        Description of digital object & The digitisation was made through TLS and Photogrammetry. The geometrical survey data consists of aligned Point clouds before and during restoration. The processed data consists of 3D NURBS Models in LOD 400 and in a variety of 3D Formats. \\
        \hhline{>{\arrayrulecolor{ruleGray}}--}
        Type of digital object & 3D NURBS Model, Photogrammetric point cloud and TLS point cloud \\
        \hhline{>{\arrayrulecolor{ruleGray}}--}
        Creator of the model & UNESCO Chair on Digital Cultural Heritage at the Cyprus University of Technology \\
        \hhline{>{\arrayrulecolor{ruleGray}}--}
        Digitisation date & May 2021 - May 2024 \\
        \hhline{>{\arrayrulecolor{ruleGray}}--}
        3D format & 3DM, E57, FBX, glTF, OBJ, STL \\
        \hhline{>{\arrayrulecolor{ruleGray}}--}
        File size & 57.5 MB \\
        \hhline{>{\arrayrulecolor{ruleGray}}--}
        URL for raw data & \url{https://datahub.egi.eu/ozw/onezone/i\#/public/shares/c5d3089cafb4a3cbe3d209f61fc7b079ch9546} \\ 
        \hhline{>{\arrayrulecolor{ruleGray}}--}
        URL for paradata & \url{https://datahub.egi.eu/share/237ea9310cfabe16d0c4aafb9857fab5ch67e0} \\ 
        \hhline{>{\arrayrulecolor{ruleGray}}--}

        \multicolumn{2}{c}{} \\[-1.5ex]

        \multicolumn{2}{@{} >{\columncolor{headerGray}} p{\dimexpr\linewidth-2\tabcolsep} @{}}{%
          \parbox[t]{\linewidth}{\vspace{0pt}\textbf{Aggregation}\vspace{2pt}\par This section contains aggregated information about all related resources pertaining to the Cultural Heritage Object.\vspace{4pt}}%
        } \\
        \hhline{>{\arrayrulecolor{ruleGray}}--}
        Aggregated CHO & \url{http://hdl.handle.net/21.15123/DKf8oUnT} \\ 
        \hhline{>{\arrayrulecolor{ruleGray}}--}
        Content provider institution & UNESCO Chair on Digital Cultural Heritage at the Cyprus University of Technology \\
        \hhline{>{\arrayrulecolor{ruleGray}}--}
        Object on provider's Website & \url{https://apsida.cut.ac.cy/items/show/49268\#?c=\&m=\&s=\&cv=\&xywh=625\%2C1392\%2C1729\%2C1044} \\ 
        \hhline{>{\arrayrulecolor{ruleGray}}--}
        Is shown by & \url{https://eureka3d.vm.fedcloud.eu/3d/00000000007EBB0A736861726547756964236564396439393833666534616133633636363764366532373063626661333766636830353633233665663637356366323134343661326366656235653866346139393233333165636865663265233832376239623936373937643263366434363161376134343337303032383661636832373733} \\ 
        \hhline{>{\arrayrulecolor{ruleGray}}--}
        Representative image & \url{https://datahub.egi.eu/api/v3/onezone/shares/data/00000000007EA8BA736861726547756964236666633635616165316232353231346238376437646137383733376161393035636830353633233665663637356366323134343661326366656235653866346139393233333165636865663265236164646431303831353864373965643965653130623866626464653332326664636832353534/content} \\ 
        \hhline{>{\arrayrulecolor{ruleGray}}--}
        Name of organisation uploading the data & Photoconsortium \\
        \hhline{>{\arrayrulecolor{ruleGray}}--}
        Copyright licence URL of the digital object & \url{http://creativecommons.org/licenses/by-sa/4.0/} \\ 
        \hhline{>{\arrayrulecolor{ruleGray}}--}

        \end{tabularx}
    } %

    \caption{The published CHO of the Lambousa Fishing Trawler as an EDM record with references to the metadata, paradata, and the actual 3D model.}
    \label{tab:europeana_data_model_lambusa}

\end{table*} %

\subsubsection{Historical Background}
Designated as a UNESCO World Heritage Site in 1985, the 14th-century Church of the Holy Cross (Timios Stavros) located in Pelendri village, is one of ten monuments that comprise the World Heritage List of Painted Churches in the Troodos Region of Cyprus. Originally, the church was a single-aisled domed which was constructed around the mid-12th century and may have functioned as the church for a cemetery. Its current appearance is the outcome of numerous additions and modifications made over different periods, with only the original apse remaining intact. An inscription found in the apse indicates that the original wall paintings date back to 1171/1172, with remnants of the decoration preserved beneath the layer of 14th-century frescoes. 

\subsubsection{Digitisation Rationale}
The UNESCO Chair on Digital Cultural Heritage at the Cyprus University of Technology’s Digital Heritage Research Lab and its collaborators took on the task of documenting the monument as a component of a collaborative training initiative under the EU-funded ERA Chair MNEMOSYNE project. The initial data collected during the 2023 mission has been refined to create a Heritage Building Information Modelling (HBIM) reference model that assists conservators and architects in the maintenance and protection of the building as part of the EUreka3D project. 

\subsubsection{Digitisation Process}
\begin{figure}
\centering
\includegraphics[width=0.5\textwidth]{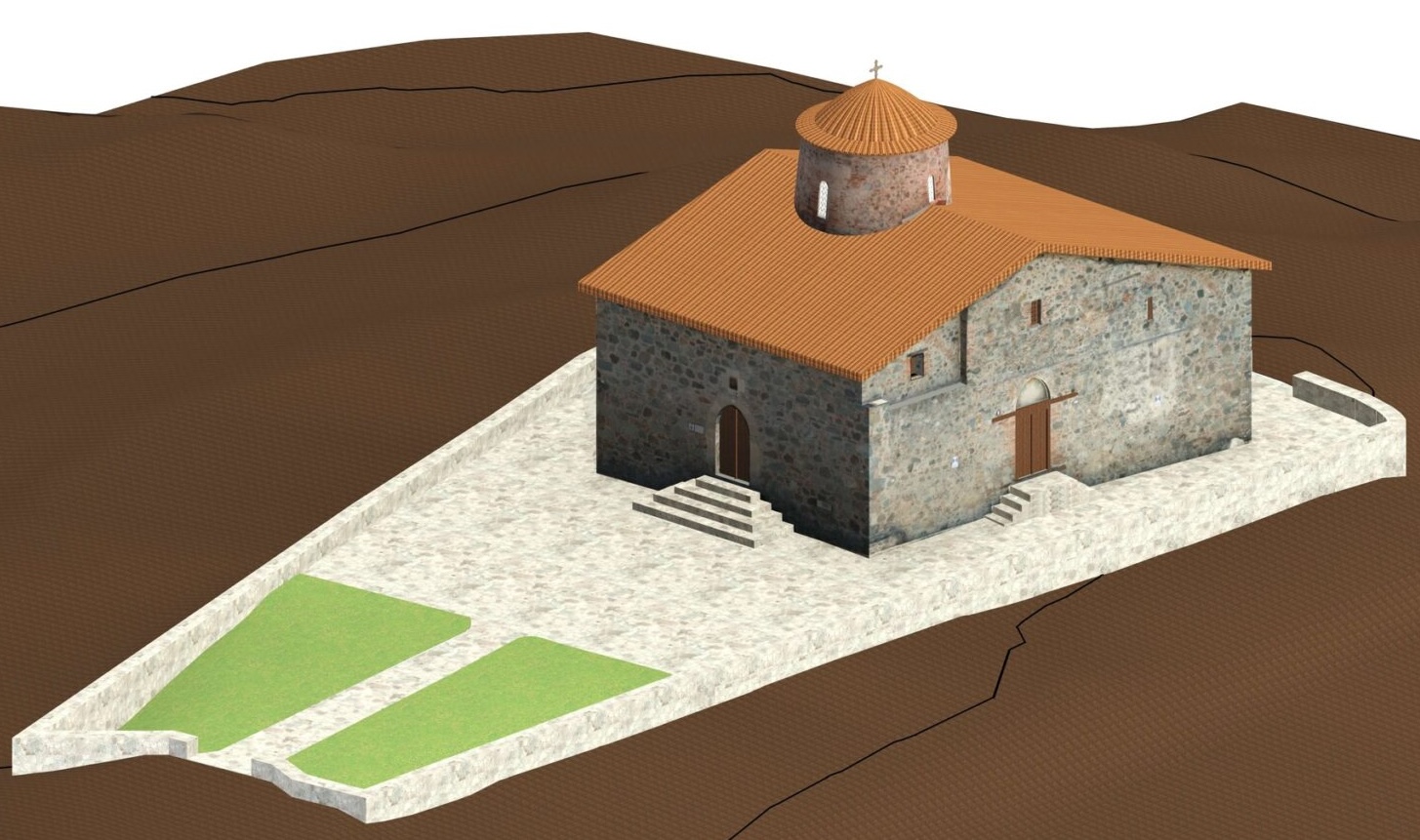}
\caption{3D model of the \textit{Church of the Holy Cross.}}
\label{fig:church}
\end{figure}
A holistic digitisation was done for the Holy Cross Church according to the VIGIE 2020/654 Study, which includes paradata, metadata, and 2D/3D data. This includes its internal and external survey through TLS and photogrammetry, the creation of an HBIM LOD 400 model, and textured mesh models of both the interior and exterior. The HBIM model created includes material and structural details of all the components of this heritage building, with the inclusion of architectural drawings that can further be used for research by architects, civil engineers, archaeologists, and historians. The texture mesh model of the exterior includes also a topographic survey, which provides details regarding the building and its surrounding mountainous landscape. The interior textured mesh model is also important because it indicates all the wall paintings in high resolution.  
The Photogrammetry and Terrestrial Laser Scanning (TLS) survey was done on 25/05/2023. The image capturing was done using DJI Phantom 4 Pro for the Exterior and Sony A7iii for the Interior. Furthermore, the Leica Viva GS15 GPS and the Leica FlexLine TS06 plus total station were used for georeferencing the point cloud model. For the processing of 158 exterior images and 1463 internal images, Reality Capture software is used. The external photogrammetry includes a textured mesh model with 27,820,560 faces, and the internal photogrammetry includes 471,200,00 faces. The TLS was done using Z+F IMAGER 5016 with 27 scan positions internally and externally, with an outcome of 88,290,000 points. Autodesk Recap was used for the Registration and Alignment of photogrammetry with TLS georeferenced point cloud. Moreover, Autodesk Revit was used for the creation of an HBIM model according to the aligned point cloud. The file formats generated are E57 and RCP for the point cloud and RVT and IFC for the BIM model. Furthermore, OBJ is exported for the online visualisation and STL for 3D printing. The digitisation and modelling results as well as paradata and metadata, were aggregated into Europeana through the EUreka3D infrastructure (3D model of the church, see Fig.~\ref{fig:church}).

\subsection{Models Processing and Results}
Each digitisation process yielded the 3D model (eg. see Fig.~\ref{fig:lambousa-model}) of the \textit{Lambousa Fishing Trawler} and accompanying files (of multiple data formats, eg.: \texttt{e57}, \texttt{obj}, \texttt{jpg}, \texttt{stl}, \texttt{3dm}) and was enriched with metadata detailed information about the object itself and vast amount of paradata describing software and hardware that was used for the digitalisation, the weather conditions and much more.

Thanks to the EGI Check-in support for eduGAIN network, the content provider could easily authenticate and access the Onedata platform. For each object, the content provider executed the templating AW, prepared the directory structure, and uploaded the results of the digitisation process.

The data verification process especially designed to handle file types resulting from the boat digitalisation validated their correctness and extracted their metadata, ingesting it into Onedata and attaching it to the uploaded files. The metadata and paradata of the 3D model were validated against the EDM scheme, and the 3D model files were annotated with it. The result were the CHOs that can be managed by the Onedata platform.

For each object, dedicated 3D processing AW converted the original \texttt{obj} files into multiple \texttt{glb} files (compatible with web-based 3D Viewer) derivatives of varying complexity and size, resulting in derivatives of the original CHO. The dedicated AW facilitated the same version of the Blender software used by the CH team, and its configuration was based on the parameters used by the CH team during the first manual post-processing phase. Thanks to running the Blender in an \textit{OCI} container, it was possible to execute the model transformation AW in a batch job manner, allowing for automation of the process. The results of the workflow processing phase are presented in Tables \ref{tab:lambousa_data}, \ref{tab:shm_data}, and  \ref{tab:church_data} show different ranges of data size and model complexity depending on supplied blender parameters. The workflows were executed on an Onedata-managed Kubernetes cluster of 4 virtual machines 4VCPU, and 16GB of RAM each.

The next step of the AW resulted in the creation of public unlisted shares for each CHO, allowing access to the 3D models from the EUreka3D 3D Viewer. 

Finally, the CHO of the EDM was created (using the ingested CHO metadata from the Onedata system) and manually validated by the content provider. The example result for the \textit{Lambousa Fishing Trawler} is depicted in Table~\ref{tab:europeana_data_model_lambusa}. The published models were then registered in the Europeana database\footnote{https://pro.europeana.eu/post/discover-how-the-eureka3d-project-supports-3d-in-the-data-space-for-cultural-heritage}. As a result, the 3D Viewer with the live 3D models of each of the CHOs, the metadata, and the paradata were embedded in Europeana, using the EDM scheme, and became available to the public. 

The presented process remains highly scalable, as the Onedata data management platform can be natively deployed on Kubernetes clusters in cloud-native environments~\cite{ppam2022}, and the AW can leverage the auto-scaling capabilities of Kubernetes~\cite{orzechowski2024cloudnativescientificworkflowmanagement}. Furthermore, the integration of the workflow engine into Onedata allows for the design of highly parallel data processing AW which execution can be delegated in parallel to any number of Kubernetes clusters.

\section{Conclusions}
\label{section:conclusions}
The linking of historical sources of different genres, their digitised data, digital research artefacts, results, and associated metadata, paradata, has been the focus of numerous projects. The recent rapid development of technologies allowing for the detailed 3D digitalisation of the CHO has presented new challenges in the standardisation and complexity management of the CHO. 
This creates challenges for existing data management systems as they need to improve metadata management capabilities to develop processes for leveraging metadata to provide better data management and semantic insight into data interconnectivity --- making the metadata a first-class component in the data management system. The application of this study in the EUreka3D project further illustrates to what level a data management system has to be enhanced for it to become a viable tool for data and metadata management for specific life-science communities.
Furthermore, the part of the EUreka3D effort was the development of a paradata model, which allows for quantitative description and qualitative comparison of digitized artefacts and the digitized process itself. The presented solution builds upon this work, exposing metadata parts of this model~\cite{marinos-study} using EDM and exposing paradata using references embedded in the CHO EMD description. Thus, the presented system, with the help of the Europeana platform, exposes a quality model of the digitized CHOs.

Because of that, we propose a flexible system of automated workflows in which logical architecture is tightly integrated with the data management system, and at the same time, its execution layer can leverage any type of job processing backend. Such a system can be used to design dedicated workflows for handling each step of CHO processing. Since the automated workflow system uses FaaS for executing individual tasks, the solution remains highly scalable when deployed on the cloud. What can foster the creation of workflow libraries that can be reused when composing dedicated workflows for different types of CHO. Moreover, such workflows are implemented in a declarative manner and combined with archiving and versioning capabilities of the data management system, allowing the reproducibility of the whole CHO processing workflow. The versioning capability of the data management system could serve as a basis for designing a system for the management of 4D CHO with an integrated 4D Viewer, allowing one to witness the evolution of 3D CHO over time. 

Our solution is characterized by: 1) the integrated management and processing of metadata, paradata, and data, 2) the flexibility and adaptability of the created AWs for different CHO processing, and 3) scalability and parallel AW execution thanks to Kubernetes integration. With that, we achieve easy
adaptation of the user coming from the cultural heritage area. 
The current approach for publishing the CHOs to Europeana, having already a European dimension, could also be applied to further disseminate the resulting 3D models to the EOSC platform.

\section*{Acknowledgements}{\small \noindent  This work is co-financed in part by program entitled \textit{International Co-financed Projects} (projects no.  5398/DIGITAL/2023/2 and 5399/DIGITAL/2023/2). JK and RGS are grateful for the support from the subvention of the Polish Ministry of Science and Higher Education assigned to AGH University of Krakow.
}\bibliographystyle{elsarticle-num} 

\bibliography{CP057}

\end{document}